\documentclass{PoS}
\usepackage{amsmath}
\usepackage{caption}
\usepackage{subcaption}
\usepackage{placeins}

\title{The Stochastic Feynman-Hellmann Method}

\ShortTitle{The Stochastic Feynman-Hellmann Method}

\author{\speaker{Arjun Singh Gambhir}, David Brantley, Pavlos Vranas\\
        Nuclear and Chemical Sciences Division, Lawrence Livermore National Laboratory\\
        E-mail: \email{gambhir1@llnl.gov}, \email{brantley3@llnl.gov}, \email{vranas2@llnl.gov}}

\author{Evan Berkowitz\\
        Institut fur Kernphysik and Institute for Advanced Simulation, Forschungszentrum Julich\\
        E-mail: \email{e.berkowitz@fz-juelich.de}}
    
    \author{Chia Cheng Chang\\
    	Interdisciplinary Theoretical and Mathematical Sciences Program
    	(iTHEMS), RIKEN 2-1 Hirosawa\\
    	E-mail: \email{ChiaChang@lbl.gov}}
    
    \author{M. A. Clark\\
        	NVIDIA Corporation\\
        	E-mail: \email{mclark@nvidia.com}}
    
    \author{Thorsten Kurth, Andr\'e Walker-Loud\\
    	National Energy Research Scientific Computing Center/Nuclear Science Division, Lawrence Berkeley National Laboratory\\
    	E-mail: \email{tkurth@lbl.gov}, \email{awalker-loud@lbl.gov}}
    
    \author{Chris Monahan\\
    	Institute for Nuclear Theory, University of Washington\\
    	E-mail: \email{cjm373@uw.edu}}
    
    \author{Amy Nicholson\\
    	Department of Physics and Astronomy, University of North Carolina\\
    	E-mail: \email{annichol@email.unc.edu}}
    
\abstract{The Feynman-Hellmann method, as implemented by Bouchard et al. [1612.06963], was recently employed successfully to determine the nucleon axial charge. A limitation of the method was the restriction to a single operator and a single momentum during the computation of each "Feynman-Hellmann" propagator. By using stochastic techniques to estimate the all-to-all propagator, we relax this constraint and demonstrate the successful implementation of this new method. We show reproduction of the axial charge on a test ensemble and non-zero momentum transfer points of the axial and vector form factors.}

\FullConference{The 36th Annual International Symposium on Lattice Field Theory - LATTICE2018\\
		22-28 July, 2018\\
		Michigan State University, East Lansing, Michigan, USA.}

\begin{document}

\section{Introduction}

The Feynman-Hellmann (FH) method connects matrix elements to variations in the spectrum. 
\begin{equation}
\frac{\partial E_n}{\partial \lambda} = <n|H_\lambda|n>
\label{FH_eq}
\end{equation}
Above, $H$ is some perturbing Hamiltonian: $H=H_0+\lambda H_\lambda$ and $|n>$ is an eigenstate of $H_0$. 
A few variations of the FH method have been employed in Lattice QCD to compute hadronic matrix elements~\cite{Chambers:2014qaa, Chambers:2015bka, Savage:2016kon,Berkowitz:2017gql,Chang:2018uxx}.
As pointed out in Ref.~\cite{Bouchard:2016heu}, these FH inspired methods are very similar to, or practically the same as those found in older publications~\cite{Maiani:1987by, Gusken:1989ad, Bulava:2011yz, deDivitiis:2012vs} and they can be rigorously connected to functional derivatives of the partition function through an application of the FH Theorem.

In the adaptation from \cite{Bouchard:2016heu}, a two-point correlation function in the presence of an external field is considered
\begin{eqnarray}
C_\lambda(t)=<\lambda|\mathcal{O}(t)\mathcal{O}^\dagger(0)|\lambda>&=&\frac{1}{\mathcal{Z}_\lambda}\int \mathcal{D}
\psi\mathcal{D}\bar{\psi}\mathcal{D}A_\mu e^{-(S+S_\lambda)}\mathcal{O}(t)\mathcal{O}(0) 
\\
\mathcal{Z}_\lambda&=&\int \mathcal{D}
\psi\mathcal{D}\bar{\psi}\mathcal{D}A_\mu e^{-(S+S_\lambda)} \nonumber \\
S_\lambda&=&\lambda\int d^4xj_\mu(x),
\end{eqnarray}
where $j_\mu(x)$ is some current density. 
\begin{equation}
\lambda j_\mu(x)=\bar{q}(x)\gamma_{\mu}\gamma_5q(x)
\end{equation}
The partial derivative of $C_\lambda(t)$ yields an expression with two terms. 
\begin{equation}
\frac{\partial C_\lambda(t)}{\partial \lambda}=
-\left[\frac{\partial_\lambda \mathcal{Z}_\lambda}{\mathcal{Z}_\lambda}
C_\lambda(t)+\frac{1}{\mathcal{Z}_\lambda}\int \mathcal{D}
\psi\mathcal{D}\bar{\psi}\mathcal{D}A_\mu 
e^{-(S+S_\lambda)}\int d^4x'j(x')\mathcal{O}(t)
\mathcal{O}(0)\right]
\label{partial_eq}
\end{equation}
The first is the vacuum matrix element, which is zero unless $j(x)$ has vacuum quantum numbers. The second term is an integral over all spacetime of the creation/annihilation operators and current. Setting $\lambda=0$ reduces $\mathcal{Z}_\lambda$ to the ordinary partition function in equation \ref{partial_eq}. Therefore, this method is employed in practice on an ensemble sampled with $\lambda=0$.
\begin{equation}
\frac{\partial C_\lambda(t)}{\partial \lambda}|_{\lambda=0}=
C_\lambda(t)\int d^4x'<\Omega|j(x')|\Omega>
-\int dt'<\Omega|T\left[\mathcal{O}(t)J(t')
\mathcal{O}(0)\right]|\Omega>
\end{equation}
Above, $J(t')$ is a time-dependent current given by $J(t')=\int d^3x' \ j(t',\overrightarrow{x}')
$. The region for which $0<t'<t$, the second term is the hadronic matrix element of interest.

In analogy to the effective mass from a standard two-point function, the long-time limit of the partial derivative of the effective mass plateaus to the matrix element of interest. 
\begin{equation}
\frac{\partial m_{\operatorname{eff}}(t,\tau)}{\partial \lambda}|_{\lambda=0}
=\frac{1}{\tau}\left[
\frac{-\partial_\lambda C_\lambda(t+\tau)}{C(t+\tau)}
-\frac{-\partial_\lambda C_\lambda(t)}{C(t)}
\right]\overrightarrow{t\rightarrow \infty}
\frac{J_{00}}{2E_0}
\label{eff_mass}
\end{equation}
There is a beautiful symmetry between equations \ref{FH_eq} and \ref{eff_mass}, hence the name, FH method. By studying $J(t')$ in all regions, including where the current is inserted outside of the hadron, a functional fitting form can be derived for which the matrix element of interest is uniquely, linearly temporally-dependent. For a rigorous treatment of this derivation, we refer the reader to \cite{Bouchard:2016heu}.  

In practice, the derivative of the correlation function is computed by applying a current globally in spacetime to a quark propagator through a matrix product. The resulting object is then treated as a source and used as a right-hand side in solving the Dirac equation.
\begin{equation}
S_{\operatorname{\Gamma}}=\sum_zS(y,z)\Gamma(z)S(z,x)
\end{equation}
We refer to $S_{\operatorname{\Gamma}}$ as the ``Feynman-Hellman" propagator. In order to compute $\partial_\lambda C_\lambda$ the FH propagator intercepts ordinary quark lines in a hadron's two-point function. This method, as numerically described above, was efficaciously used to calculate the axial coupling to within $1\%$  uncertainty \cite{Chang:2018uxx}.

There are several benefits to using this technique for computing correlation functions. Since the insertion time is summed over, there is only one time variable that may be varied, leading to potentially easier systematics to control. Additionally, for the cost of a single propagator, information from all timeslices is gained. Furthermore, $S_{\operatorname{\Gamma}}$ is reusable for a variety of different sink interpolators, including different choices of sink smearing and different hadrons altogether.

\section{Stochastic Feynman-Hellmann}

The FH method has a number of disadvantages compared to the more conventional sequential source approach to computing three-point functions \cite{b1986lattice,Bratt:2010jn}. The current time is summed everywhere on the lattice, including over sites outside of the hadron and at the source/sink, so called contact terms. This yields extra terms that do not appear in other methods. In addition, explicit time dependence of the current is lost. This complicates the analysis for non-zero momentum transfer points, in which transition amplitudes appear due to differing initial and final state energies. Finally, each choice of $\Gamma$ and momentum injection costs an individual FH propagator, resulting in high computational costs to calculate form factors.

The aim of this work is to remedy two out of the three disadvantages. To achieve this, we employ a stochastic basis, whose vectors obey the property: $<\eta(i)\eta^*(j)>=\delta_{ij}$. If we consider one spin/color component of a propagator, the Dirac equation is solved as
\begin{eqnarray*}
D|\psi>&=&|\chi> \\
|\psi>&=&D^{-1}|\chi>, \\ 
\end{eqnarray*}
where $|\chi>$ is a spin/color component of the source and $|\psi>$ is of the propagator. A component of the FH propagator is 
\begin{equation}
|\phi>=D^{-1}\Gamma e^{iq\cdot x}|\psi>.
\end{equation}
Inserting an outer product of noise vectors factorizes the algorithm, allowing different matrix elements and momentum-transfer points to be computed without requiring additional solves. 
\begin{equation}
|\phi '>=\frac{1}{N}\sum_{i=1}^{N}D^{-1}|\eta_i><\eta_i|\Gamma e^{iq\cdot x}|\psi>
\end{equation}

There are many possible choices for the type of noise basis, such as Gaussian, $\mathbb{Z}_2$, $\mathbb{Z}_4$, or some kind of dilution pattern. In the context of disconnected diagrams or distillation, the effectiveness of different types of noise and variance reduction techniques has been studied extensively \cite{Hutchinson_90, Dong:1993pk, Neff:2001zr, Foley:2005ac, Babich:2007jg, Morningstar_Peardon_etal_2011, Stathopoulos:2013aci, Endress:2014qpa, Gambhir:2016uwp, Gambhir:2016jul}. For this work we employ fully spin and color diluted hierarchical probing \cite{Stathopoulos:2013aci}.

\section{Numerical Results}
\begin{figure}[!ht]
\centering

\subcaptionbox{$g_A$ with $Q^2=0$\label{sFH_q0}}
{\includegraphics[width=0.45\textwidth]{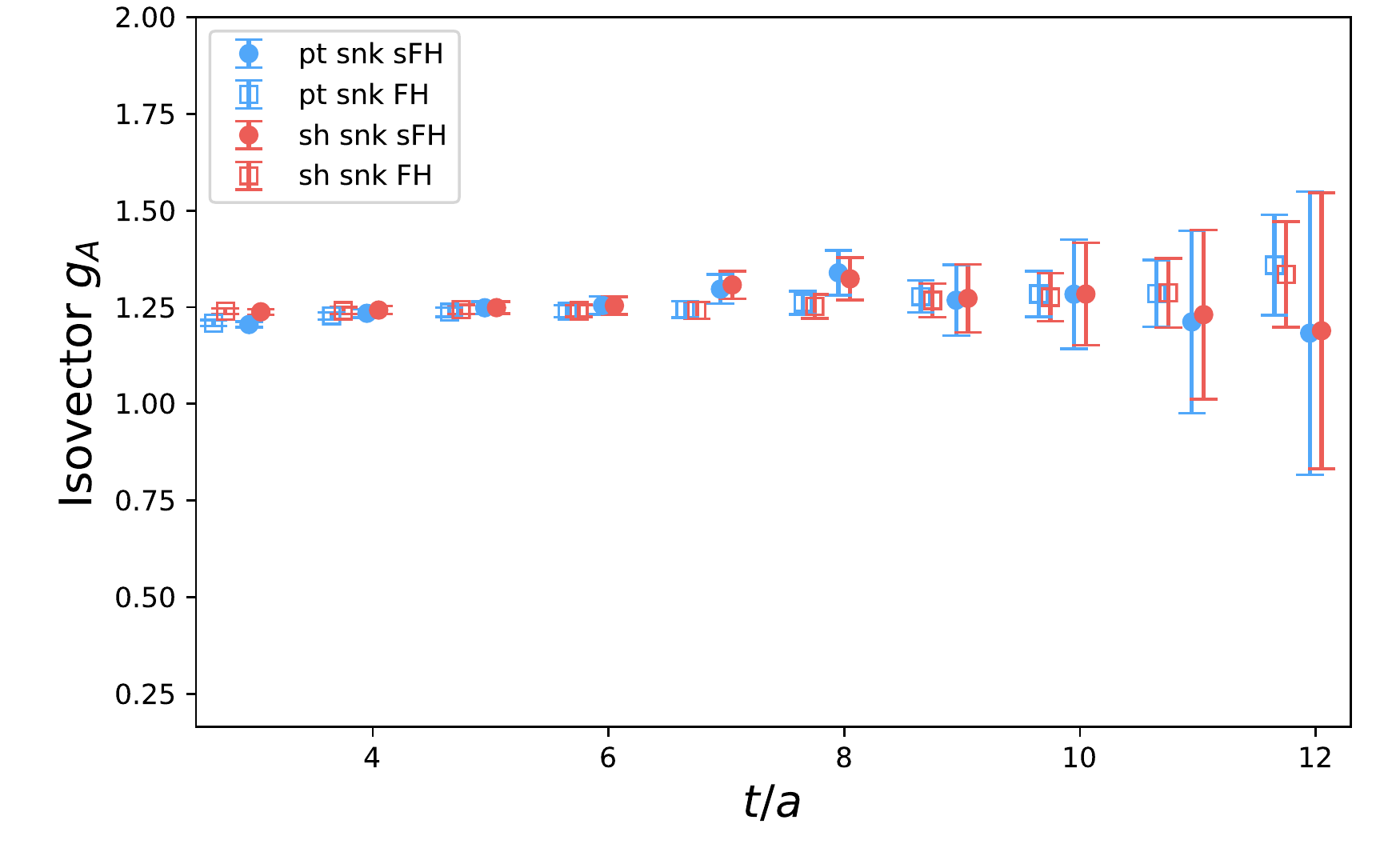}}
\subcaptionbox{$g_V$ with $Q^2=0$\label{sFH_gV_q0}}
{\includegraphics[width=0.45\textwidth]{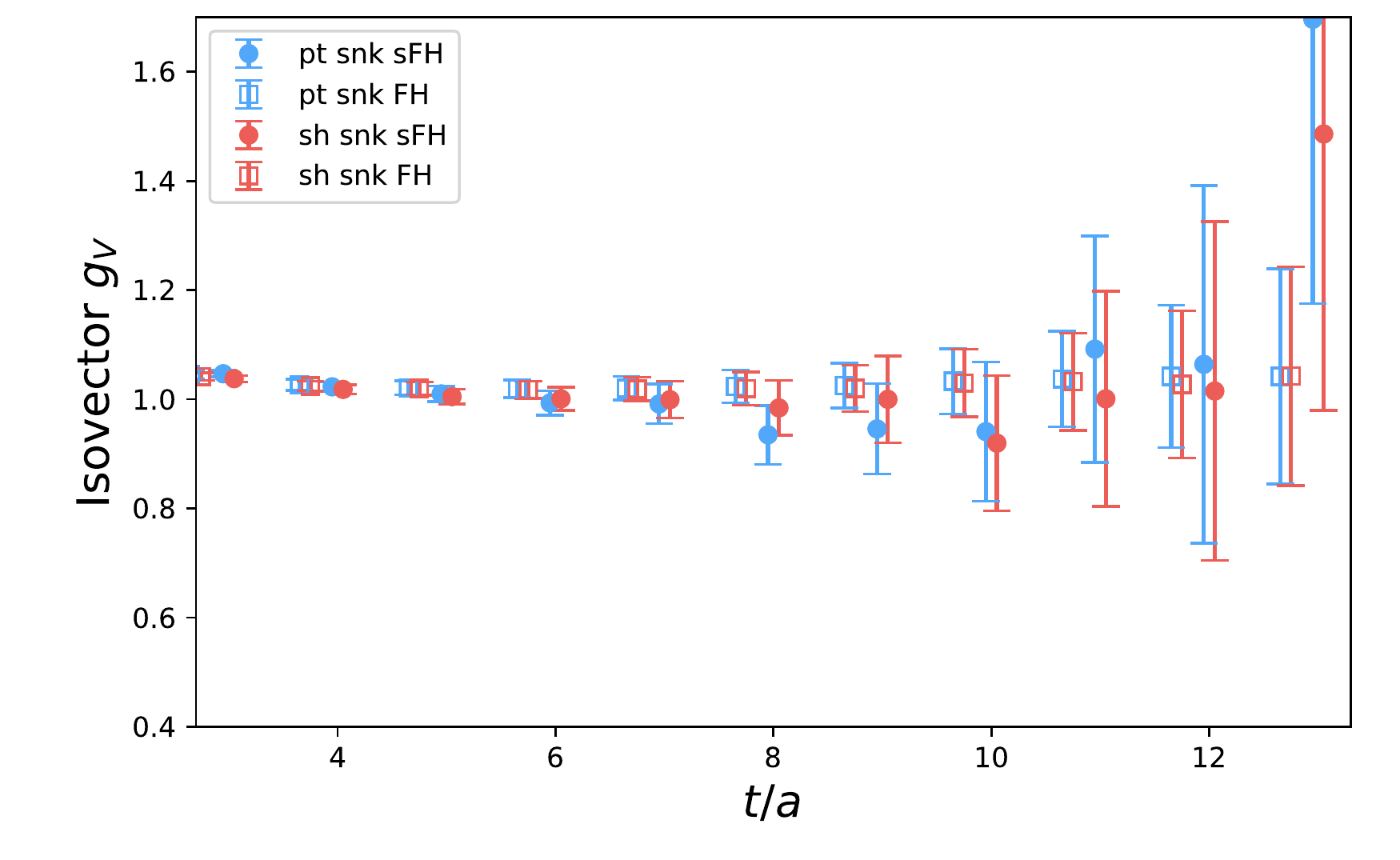}}
\caption{Isovector $g_A$ and $g_V$ with regular FH and stochastic FH (sFH) are plotted. Both methods are at comparable statistics with about 1000 gauge field configurations and 8 smeared sources. Both point (pt) and smeared (sh) sinks are constructed. For the stochastic case, 32 hierarchical probing vectors are used.}
\end{figure}

\begin{figure}[h]
\centering
\includegraphics[width=0.6\textwidth]{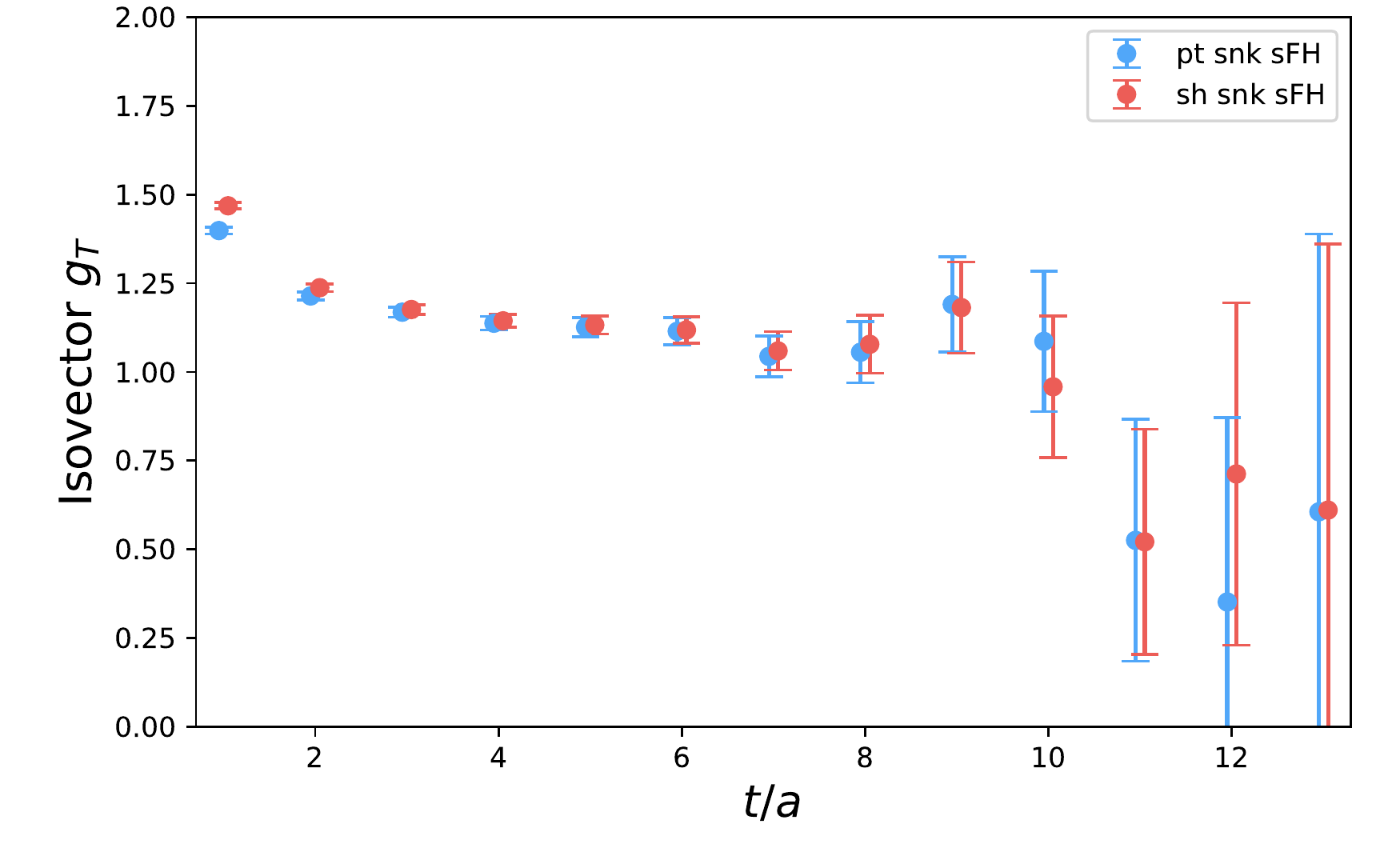}
\caption{Isovector $g_T$ with the stochastic method is shown. This calculation was done with the same stochastic basis as previous figures, with identical statistics of roughly 1000 configurations and 8 sources.}
\label{sFH_T12_q0}
\end{figure}

\begin{figure}[!ht]
\centering

\subcaptionbox{$g_A$ with $Q^2=.18 \ \operatorname{GeV}^2$\label{sFH_q1}}
{\includegraphics[width=0.45\textwidth]{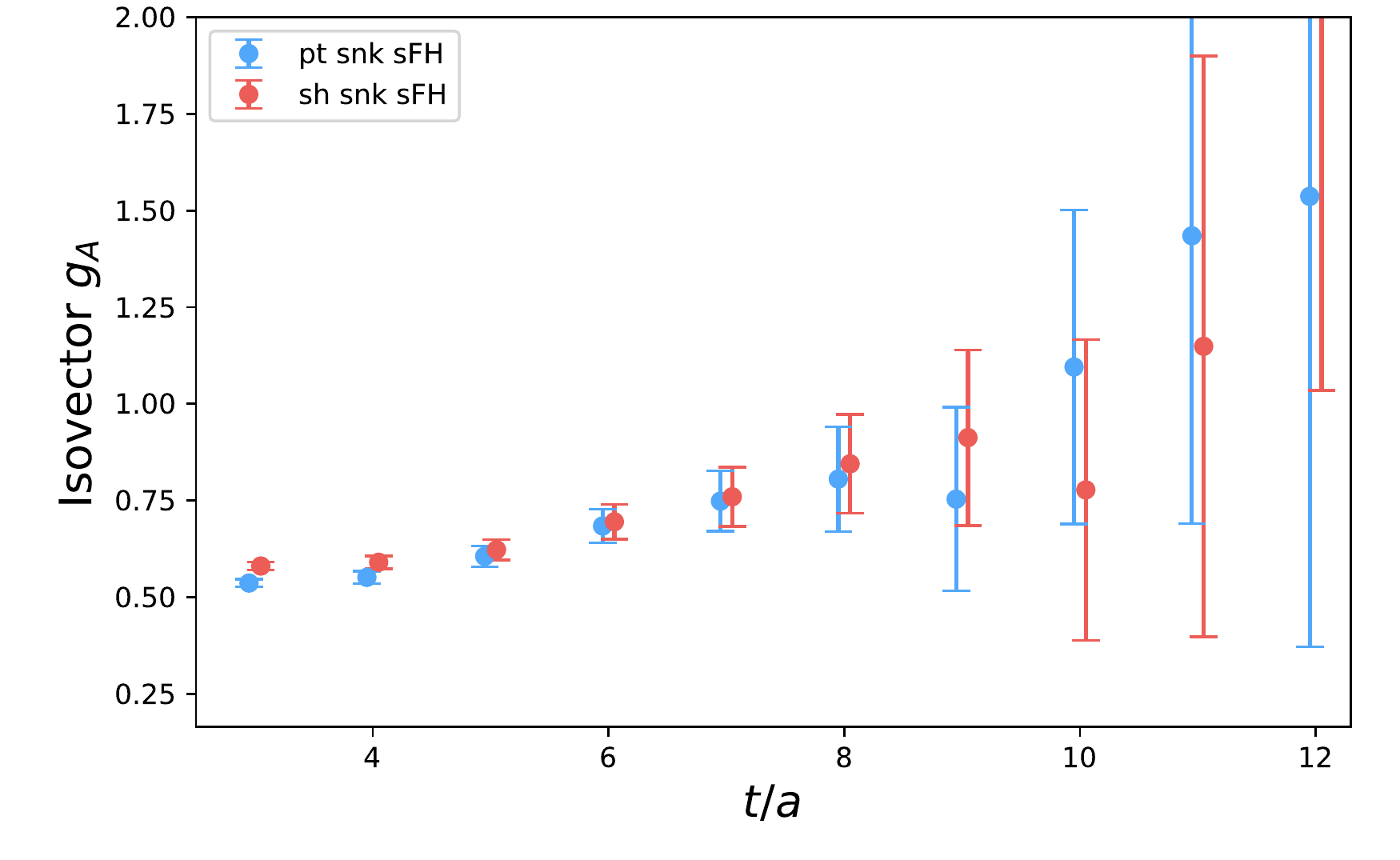}}
\subcaptionbox{$g_V$ with $Q^2=.18 \ \operatorname{GeV}^2$\label{sFH_gV_q1}}
{\includegraphics[width=0.45\textwidth]{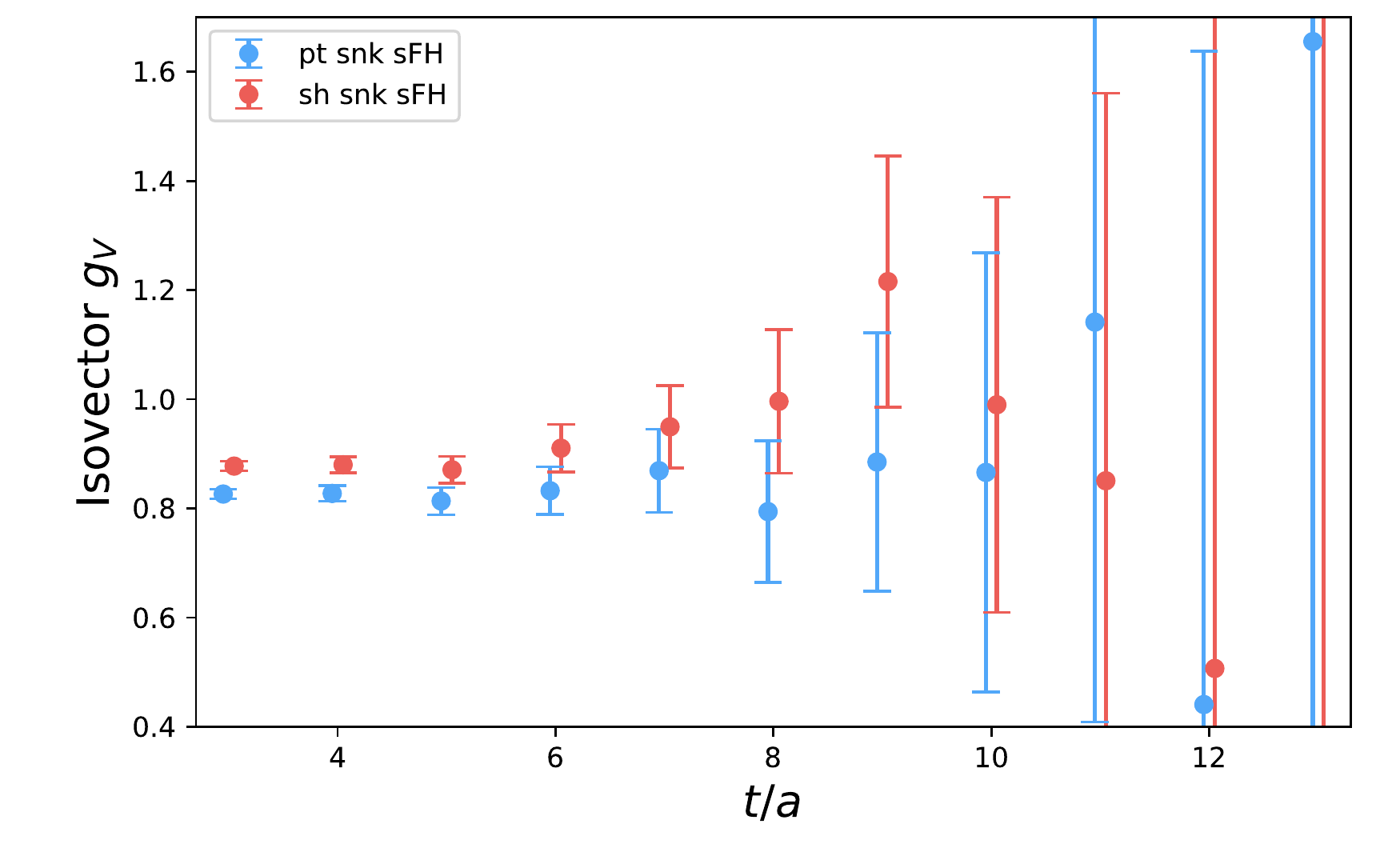}}
\caption{Isovector $g_A$ and $g_V$ with one unit of lattice momentum are shown. This is with approximately 1000 configurations, 8 sources, and 32 hierarchical probing vectors.}
\end{figure}

\FloatBarrier
The stochastic FH method is tested on a M\"{o}bius Domain Wall on gradient flowed HISQ mixed action. The lattice spacing, pion mass, and volume of the ensemble are $a=.12 \ \operatorname{fm}$, $m_\pi=310 \ \operatorname{MeV}$, and $m_\pi L=4.5$ respectively. Details of the action setup and ensemble can be found in \cite{Berkowitz:2017opd}. Figures \ref{sFH_q0} and \ref{sFH_gV_q0} show the calculation of $g_A$ and $g_V$ with regular and stochastic FH methods. Figure \ref{sFH_T12_q0}, \ref{sFH_q1}, and \ref{sFH_gV_q1} demonstrate the computation of other observables by reusing the same stochastic basis. 

In these figures, the derivative of the effective mass is plotted by constructing the correlated difference defined in equation \ref{eff_mass}. In the long-time limit, this quantity should plateau to the matrix element. From Figures \ref{sFH_q0} and \ref{sFH_gV_q0}, it's clear that the stochastic basis successfully reproduces the ``exact" answer, albeit with larger statistical uncertainty. In the case of Figure \ref{sFH_T12_q0}, there is no exact calculation to compare to, however the signal looks clean and plateaus to a reasonable value of the tensor charge \cite{Gupta:2018qil}.

Figures \ref{sFH_q1} and \ref{sFH_gV_q1} illustrate $g_A$ and $gV$ with one unit of lattice momentum. This is done by fixing the sink and current with one unit of momentum in the same direction. It should be noted however that with the stochastic FH method, arbitrary momenta may be chosen for both the current insertion and sink. An analysis that takes advantage of different momentum combinations at the current and sink is forthcoming. The statistical uncertainty for non-zero momentum is large; enhancements to the basis are also currently being explored.
 
\FloatBarrier

\section{Conclusion and Outlook}

We study an algorithm that computes the FH propagator with a stochastic basis, extending the original FH method to give arbitrary observables and non-zero momentum transfer points in a feasible amount of computing time. This is achieved through a stochastic noise basis. It should be noted that this basis is required to compute disconnected diagrams, therefore amortizing the cost for the connected calculation. Additionally, the current insertion time may be exposed by introducing a delta function in time to the bilinear, thereby recovering the full dependence of the three-point function. Finally, it has been shown in \cite{Gambhir:2016uwp, Gambhir:2016jul} that singular value deflation plays a synergistic role with hierarchical probing in greatly reducing variance. Incorporating deflation into this method is being explored.

\section*{Acknowledgments}

This work was supported by an award of computer time by the Lawrence Livermore National Laboratory (LLNL) Multiprogrammatic and Institutional Computing program through a Tier 1 Grand Challenge award. We would like to thank Balint Joo for general support with USQCD software and the larger CalLat collaboration for useful discussions. The work of ASG was performed under the auspices of the U.S. Department of Energy by LLNL under Contract No. DE-AC52-07NA27344. This research used the Sierra computer operated by the Lawrence Livermore National Laboratory for the Office of Advanced Simulation and Computing and Institutional Research and Development, NNSA Defense Programs within the U.S. Department of Energy.

\bibliographystyle{JHEP}
\bibliography{qcd}  

\end{document}